\documentclass[12pt]{article}%
\usepackage{amsmath}
\usepackage{amsfonts}
\usepackage{amssymb}
\usepackage{graphicx}%
\providecommand{\U}[1]{\protect\rule{.1in}{.1in}}
\begin{document}
\begin{titlepage}
\ \\
\begin{center}
\LARGE
{\bf
Controlled Hawking Process\\
by Quantum Energy Teleportation
}
\end{center}
\ \\
\begin{center}
\large{
Masahiro Hotta
}\\
\ \\
\ \\
{\it
Department of Physics, Faculty of Science, Tohoku University,\\
Sendai 980-8578, Japan\\
hotta@tuhep.phys.tohoku.ac.jp
}
\end{center}
\begin{abstract}
In this paper, a new quantum mechanical method to extract energy from black holes
with contracting horizons is proposed. The method is based on a gedanken experiment
on quantum energy teleportation (QET), which has been recently proposed in quantum information theory.
We consider this QET protocol for N massless fields in near-horizon regions of large-mass black holes
with near-horizon geometry described by the Minkowski metric.
For each field, a two-level spin is strongly coupled with the local quantum fluctuation
outside the horizon during a short time period. After the measurement of N fields,
N-bit information is obtained. During the measurement, positive-energy wave packets of the fields
form and then fall into the black hole.
The amount of excitation energy is independent of the measurement result.
After absorption of the wave packets and increase of the black hole mass,
a measurement-result-dependent local operation of the N fields is performed outside the horizon.
Then, accompanying the extraction of positive energy from the quantum fluctuation by the operation,
negative-energy wave packets of the fields form and then fall into the black hole, decreasing the black hole mass.
This implies that a part of the absorbed positive energy emitted from the measurement devices
is effectively retrieved from the black hole via the measurement results.
\end{abstract}
\end{titlepage}

\bigskip

\section{Introduction}

\ \newline

\bigskip

Essentially, two methods are known by which energy can be extracted from black
holes. One is the classical method that is based on the existence of
ergospheres of rotating black holes. It includes the Penrose mechanism of test
particles \cite{penrose}, the superradiance of an electromagnetic field,
\cite{sr} and the Blandford-Znajek effect of accretion disks with magnetic
field \cite{bz} for Kerr black holes. Moreover, it is possible to extract
energy from Reissner-Nordstrom charged black holes because ergosphere-like
structures of the black holes appear for charged particles \cite{charged
hole}. Even though energy can be extracted from black holes using these
classical methods, the horizon areas of the black holes do not decrease. If
the quantum effect of matter is taken into account, energy extraction with a
decrease in horizon area is possible. This is the second method of energy
extraction, called the Hawking effect \cite{h}, which describes black hole
evaporation. Black holes emit thermal radiation, accompanied by generation of
negative energy flux that falls into the black holes. The Hawking effect takes
place spontaneously outside the horizon and cannot be stopped artificially. In
this paper, a new quantum mechanical method of energy extraction from black
holes is proposed. This method is based on a gedanken experiment on quantum
energy teleportation (QET), which has been recently proposed in quantum
information theory \cite{hotta1}--\cite{hotta4}. In this method, energy is
extracted by a local quantum operation after a quantum measurement of quantum
fluctuation of fields outside the horizons; thus, it is not a spontaneous
phenomenon like the Hawking process. Besides, the time scale of the QET
process is very short compared to the Hawking-process time scale of order of
inverse of black-hole temperature.

The protocols of QET consist of local operations and classical communication.
By measuring the local fluctuation induced by a zero-point oscillation in the
ground state of a many-body quantum system and by announcing the measurement
result to distant points, energy can be effectively teleported without
breaking any physical laws including causality and local energy conservation.
The key point is that there exists quantum correlation between local
fluctuations of different points in the ground state. Therefore, the
measurement results of local fluctuation in a region include information about
fluctuations in other regions. By selecting and performing a proper local
operation that is based on the announced information, the local zero-point
oscillation of a place far from the measurement point can be squeezed,
yielding negative energy density. During the local operation, surplus energy
of the local fluctuation is emitted to the external systems and can be harnessed.

We consider this QET protocol for $N$ massless fields in the near-horizon
regions of large-mass black holes with near-horizon geometry described by the
Minkowski metric. The Hartle--Hawking state of the fields in the black hole
spacetime, which is the thermal equilibrium state with the Hawking temperature
\cite{hh}, is reduced to the Minkowski vacuum state in flat spacetime. For
each field, a two-level spin is strongly coupled with the local quantum
fluctuation outside the horizon during a short time period. The spin is
measured to give 1-bit information about the fluctuation. After the
measurement of $N$ fields, $N$-bit information, denoted by $D_{k}$
($k=1\sim2^{N}$), is obtained. The post-measurement quantum states outputting
$D_{k}$ are not the vacuum states but the states excited by the measurement
devices. Positive-energy wave packets of the fields form and then fall into
the black hole. The amount of excitation energy is independent of $D_{k} $.
After absorption of the wave packets and increase in the black hole mass, a
$D_{k}$-dependent local operation of the $N~$fields is performed outside the
horizon. Then, along with the extraction of positive energy from the
fluctuation by the operation, negative-energy wave packets of the fields form
and then fall into the black hole. The mass of the black hole decreases due to
absorption of the negative flux. This means that a part of the absorbed
positive energy emitted from the measurement devices is effectively retrieved
from the black hole via the measurement results. Though energy extracted by
QET does not exhibit thermal property like Hawking radiation, the energy
extraction process involving measurement and the local operation, which
generates a pair of excitations with positive and negative energy, is regarded
as a controlled Hawking process. It is noteworthy that if a unitary operation
corresponding to a different $N$-bit sequence $D_{k^{\prime}\neq k}$ is
performed, the amount of energy extracted by QET decreases. Without any
information about the correct $D_{k}$, a unitary operation corresponding to a
random $N$-bit sequence does not extract energy from the fields, but gives
energy to the fields on average. This implies that a black hole, which
swallows the wave packets generated by the measurement with output $D_{k}$,
remembers the information about $D_{k}$. Thus, black holes absorbing quantum
states with different $D_{k}$ are physically different even though their
classical geometries are the same.

The paper is organized as follows. In section 2, we briefly review the horizon
shift of large-mass black holes in classical theory, using the CGHS model. In
section 3,\ we analyze a gedanken experiment on QET in a near-horizon region.
The summary and discussion are provided in section 4. We adopt the natural
unit $c=\hbar=1$.

\section{ Horizon Shift in Classical Gravity}

\bigskip

To briefly review the property of horizon shift in classical gravity, we adopt
a two-dimensional soluble dilaton gravity model called the CGHS model
\cite{CGHS} for simplicity. For higher-dimensional spherically symmetric black
holes with large masses, we obtain essentially the same conclusion by treating
falling matter with a spherically symmetric distribution. In the CGHS model,
the action is given by
\[
S=\int d^{2}x\sqrt{-g}\left(
\begin{array}
[c]{c}%
e^{-2\varphi}\left(  R+4(\nabla\varphi)^{2}+4\Lambda^{2}\right) \\
-\frac{1}{2}\sum_{n=1}^{N}(\nabla f_{n})^{2}%
\end{array}
\right)  ,
\]
where $\varphi$ is a real scalar dilaton field, $\Lambda$ is a positive
constant, and $f_{n}$ denotes $N$ real massless matter fields. The static
black hole solutions are given by%

\[
ds^{2}=-\left(  1-\frac{M}{\Lambda}e^{-2\Lambda r}\right)  d\tau^{2}+\left(
1-\frac{M}{\Lambda}e^{-2\Lambda r}\right)  ^{-1}dr^{2},
\]
where $M$ represents the black hole mass. The horizon radius is given by
$r_{H}=\frac{1}{2\Lambda}\ln\left(  \frac{M}{\Lambda}\right)  $. Using new
coordinates defined by $x^{\pm}=\pm\frac{1}{\sqrt{M\Lambda}}\sqrt{e^{2\Lambda
r}-\frac{M}{\Lambda}}e^{\pm\Lambda\tau}$, the metric is rewritten as%
\[
ds^{2}=-\frac{dx^{+}dx^{-}}{1-\Lambda^{2}x^{+}x^{-}}.
\]
The future horizon stays at $x^{-}=0$ in the new coordinates. Because the
scalar curvature on the horizon is proportional to $\Lambda^{2}$, a very small
$\Lambda$ yields the Minkowski spacetime as the near-horizon geometry. In the
near-horizon region with $\left\vert x^{\pm}\right\vert \ll1/\Lambda$, the
metric is reduced to $ds^{2}=-dx^{+}dx^{-}$. The matter field $f_{n}$ obeys
the equation $\partial_{+}\partial_{-}f_{n}=0$ in the region and can be
decomposed into left- and right-moving solutions as follows: $f_{n}%
=f_{Ln}(x^{+})+f_{Rn}(x^{-})$. Solutions with falling matter into the black
hole with mass $M$ are also analytically obtained \cite{CGHS} as%

\[
ds^{2}=-\frac{dx^{+}dx^{-}}{1-\Lambda^{2}x^{+}x^{-}-\frac{\Lambda}{2M}%
\int_{-\infty}^{x^{+}}dy^{+}\int_{-\infty}^{y^{+}}dz^{+}T_{++}(z^{+})},
\]
where $T_{++}(x^{+})$ is the incoming energy flux and is given by $\sum
_{n=1}^{N}\left(  \partial_{+}f_{Ln}(x^{+})\right)  ^{2}$. Therefore, the
metrics of general solutions in the $x^{\pm}$ coordinates also become the flat
metric $ds^{2}=-dx^{+}dx^{-}$ when $M\gg\Lambda$ and $\left\vert x^{\pm
}\right\vert \ll1/\Lambda$. Hence, the horizon shift due to absorption of
matter cannot be observed in these coordinates of the near-horizon region.

In order to observe the horizon shift, rescaled coordinates are available;
these are given by $X^{\pm}=\sqrt{\frac{M}{\Lambda}}x^{\pm}$ \cite{CGHS}. The
metric is then rewritten as%

\begin{equation}
ds^{2}=-\frac{dX^{+}dX^{-}}{\frac{M}{\Lambda}-\Lambda^{2}X^{+}X^{-}-\frac
{1}{2}\int_{-\infty}^{X^{+}}dY^{+}\int_{-\infty}^{Y}dZ^{+}T_{++}(\sqrt
{\frac{\lambda}{M}}Z^{+})}. \label{e1}%
\end{equation}
From Eq. (\ref{e1}), the horizon shift is completed soon after the falling
matter passes. If the support of $T_{++}(x^{+})$ is given by $\left(
-\infty,x_{s}^{+}\right)  $, the horizon stops soon after $X_{s}^{+}%
=\sqrt{\frac{M}{\Lambda}}x_{s}^{+}$. The metric in the region with
$X^{+}>X_{s}^{+}$ is given by
\[
ds^{2}=-\frac{dX^{+}dX^{-}}{\frac{M^{\prime}}{\Lambda}-\Lambda^{2}X^{+}\left(
X^{-}-X_{H}\right)  },
\]
where $M^{\prime}$ is the enlarged black hole mass and the shifted future
horizon stays at $X^{-}=X_{H}=-\frac{1}{2\Lambda^{2}}\sqrt{\frac{M}{\Lambda}%
}\int_{-\infty}^{x_{s}^{+}}T_{++}(x^{+})dx^{+}$. Even for higher dimensional
spherically symmetric black holes of Einstein gravity, the horizon also
stabilizes soon after the falling matter passes if the matter takes the shape
of spherical shells. This is because any spherically symmetric vacuum solution
of the Einstein equation is static and equal to the Schwarzschild solution.
After the horizon is shifted, the classical geometry of the black holes holds
no information except the total energy. However, as explained in the next
section, the black hole remembers more information quantum mechanically.

\bigskip

\section{ Quantum Energy Teleportation}

\bigskip

The horizon of the initial black hole stays at $x^{-}=0$. Let us consider $N$
quantum fields $\hat{f}_{n}\,\ (n=1\sim N)$ in the near-horizon region with
the $x^{\pm}$ coordinates. Because the fields are $N$ copies of a massless
scalar field, we can concentrate on one of the fields and disregard the index
$n$ for the time being. We consider a QET process with a time scale much
shorter than inverse of the black hole mass which is the time scale of
generation of the Hawking radiation. Thus, the Hawking process can be
neglected in the later discussion. The equation of motion is written as
$\partial_{+}\partial_{-}\hat{f}=0$, and the general solution is solved by the
quantum left- and right-moving solutions as $\hat{f}=\hat{f}_{+}\left(
x^{+}\right)  +\hat{f}_{-}\left(  x^{-}\right)  $. The spacetime coordinates
of the Minkowski metric are given by $t=\frac{1}{2}\left(  x^{+}+x^{-}\right)
$ and $x=\frac{1}{2}\left(  x^{+}-x^{-}\right)  $. The canonical conjugate
momentum operator of $\hat{f}\left(  x\right)  =\hat{f}|_{t=0}$ is defined by
$\hat{\Pi}(x)=\partial_{t}\hat{f}|_{t=0}$ and it satisfies the standard
commutation relation, $\left[  \hat{f}\left(  x\right)  ,~\hat{\Pi}\left(
x^{\prime}\right)  \right]  =i\delta\left(  x-x^{\prime}\right)  $. The
left-moving wave $\hat{f}_{+}\left(  x^{+}\right)  $ can be expanded in terms
of plane-wave modes as%

\[
\hat{f}_{+}\left(  x^{+}\right)  =\int_{0}^{\infty}\frac{d\omega}{\sqrt
{4\pi\omega}}\left[  \hat{a}_{\omega}^{L}e^{-i\omega x^{+}}+\hat{a}_{\omega
}^{L\dag}e^{i\omega x^{+}}\right]  ,
\]
where $\hat{a}_{\omega}^{L}$ $(\hat{a}_{\omega}^{L\dag})$ is an annihilation
(creation) operator of a left-moving particle and satisfies%

\begin{equation}
\left[  \hat{a}_{\omega}^{L},~\hat{a}_{\omega^{\prime}}^{L\dag}\right]
=\delta\left(  \omega-\omega^{\prime}\right)  . \label{c2}%
\end{equation}
The right-moving wave $\hat{f}_{-}\left(  x^{-}\right)  $ can be also expanded
in the same way using the plane-wave modes. The energy density operator is
given by
\[
\hat{\varepsilon}(x)=\frac{1}{2}:\hat{\Pi}(x)^{2}:+\frac{1}{2}:\left(
\partial_{x}\hat{f}(x)\right)  ^{2}:,
\]
where $::$ denotes the normal order of creation-annihilation operators for the
plain-wave modes. The Hamiltonian is given by $\hat{H}=\int_{-\infty}^{\infty
}\hat{\varepsilon}(x)dx$. In the near-horizon region, the Hartle--Hawking
state \cite{hh} is described by the Minkowski vacuum state $|0\rangle$,
defined by $\hat{H}|0\rangle=0$. It is also satisfied that $\langle
0|\hat{\varepsilon}\left(  x\right)  |0\rangle=0$. Let us define the chiral
momentum operators as
\[
\hat{\Pi}_{\pm}(x)=\hat{\Pi}\left(  x\right)  \pm\partial_{x}\hat{f}(x).
\]
Then, the energy density can be rewritten as
\begin{equation}
\hat{\varepsilon}\left(  x\right)  =\frac{1}{4}:\hat{\Pi}_{+}\left(  x\right)
^{2}:+\frac{1}{4}:\hat{\Pi}_{-}\left(  x\right)  ^{2}:. \label{504}%
\end{equation}

Following the method given in \cite{hotta1}, we perform QET for the state
$|0\dot{\rangle}$ as follows. Figure 1 shows a spacetime diagram that
expresses this QET experiment. Consider a probe system $P$ of a two-level spin
located in a small compact region $\left[  x_{1},x_{2}\right]  $ satisfying
$x_{1}>0$ outside the horizon in order to detect fluctuations of $\hat{f}$. In
a similar way of particle-detector interaction of Unruh \cite{u}, we introduce
a measurement Hamiltonian between $\hat{f}$ and the spin probe such that
\[
\hat{H}_{m}(t)=g(t)\hat{G}_{+}\otimes\hat{\sigma}_{y},
\]
where $g(t)$ is a time-dependent real coupling constant, $\hat{G}_{+}$ is
given by
\begin{equation}
\hat{G}_{+}=\frac{\pi}{4}+\int_{-\infty}^{\infty}\lambda(x)\hat{\Pi}%
_{+}\left(  x\right)  dx, \label{600}%
\end{equation}
$\lambda(x)$ is a real function with support $\left[  x_{1},x_{2}\right]  $,
and $\hat{\sigma}_{y}$ is the $y$-component of the Pauli matrices of the probe
spin. We assume that the initial state of the probe is the up state
$|+\frac{1}{2}\rangle$ of the $z$-component $\hat{\sigma}_{z}$. The time
dependence of $g(t)$ is assumed to be generated by a device with an external
energy supply from spatial infinity, respecting local energy conservation. In
the later analysis, we choose a sudden switching form such that $g(t)=\delta
(t-0)$. After the interaction is switched off, we measure the z-component
$\hat{\sigma}_{z}$ for the probe spin. If the up or down state, $\left\vert
+\frac{1}{2}\right\rangle $ or $\left\vert -\frac{1}{2}\right\rangle $, of
$\hat{\sigma}_{z}$ is observed, we assign the bit number $b=0$ or $1$,
respectively, to the measurement result. The measurement is completed at
$t=+0$. The time evolution of this measurement process with output $b$ can be
described by the measurement operators $\hat{M}_{b}$, which are often used in
quantum information theory \cite{nc} and which satisfy%

\[
\hat{M}_{b}\rho\hat{M}_{b}^{\dag}=\operatorname*{Tr}_{P}\left[  \left(
I\otimes\left\vert (-1)^{b}\frac{1}{2}\right\rangle \left\langle (-1)^{b}%
\frac{1}{2}\right\vert \right)  \hat{V}\left(  \rho\otimes\left\vert +\frac
{1}{2}\right\rangle \left\langle +\frac{1}{2}\right\vert \right)  \hat
{V}^{\dag}\right]  ,
\]
where $\rho$ is an arbitrary density operator of the field, the time evolution
operator $\hat{V}=\operatorname*{T}\exp\left[  -i\int_{-0}^{t}\hat{H}%
_{m}(t^{\prime})dt^{\prime}\right]  $ is computed as $\exp\left[  -i\hat
{G}_{+}\otimes\hat{\sigma}_{y}\right]  $ for $t>0$, and the trace
$\operatorname*{Tr}_{P}$ is taken to the probe system. The measurement
operators $\hat{M}_{b}$ are evaluated as%

\[
\hat{M}_{b}=\left\langle (-1)^{b}\frac{1}{2}\right\vert \exp\left[  -i\hat
{G}_{+}\otimes\hat{\sigma}_{y}\right]  \left\vert +\frac{1}{2}\right\rangle .
\]
Hence, we obtain the explicit expression of $\hat{M}_{b}$ such that%

\begin{align}
\hat{M}_{0}  &  =\cos\hat{G}_{+},\label{401}\\
\hat{M}_{1}  &  =\sin\hat{G}_{+}. \label{402}%
\end{align}
For the vacuum state $|0\rangle$, the probability of obtaining $b$ in the
measurement $\langle0|\hat{M}_{b}^{\dag}\hat{M}_{b}|0\rangle$ is independent
of $b~$and is given by $1/2$. The explicit calculation of $\langle0|\hat
{M}_{b}^{\dag}\hat{M}_{b}|0\rangle$ is as follows.%

\begin{align*}
\langle0|\hat{M}_{b}^{\dag}\hat{M}_{b}|0\rangle &  =\langle0|\left[  \frac
{1}{2}+\frac{(-1)^{b}}{2}\cos\left(  2\hat{G}_{+}\right)  \right]  |0\rangle\\
&  =\frac{1}{2}-\frac{(-1)^{b}}{2}\langle0|\sin\left(  2\int_{-\infty}%
^{\infty}\lambda(x)\hat{\Pi}_{+}\left(  x\right)  dx\right)  |0\rangle\\
&  =\frac{1}{2}.
\end{align*}
In the above calculation, $\sin\left(  2\int_{-\infty}^{\infty}\lambda
(x)\hat{\Pi}_{+}\left(  x\right)  dx\right)  $ is calculated as the sum of the
products of odd-number field operators and the correlation functions with
odd-number field operators vanish for the vacuum state $|0\rangle$ of this
free field. The post-measurement states of $\hat{f}$ for the result $b$ are
calculated as%

\begin{equation}
|\psi_{b}\rangle=\sqrt{2}\hat{M}_{b}|0\rangle=\frac{i^{b}}{\sqrt{2}}\left(
e^{-\frac{\pi}{4}i}|\lambda\rangle+\left(  -1\right)  ^{b}e^{\frac{\pi}{4}%
i}|-\lambda\rangle\right)  , \label{ps}%
\end{equation}
where $|\pm\lambda\rangle$ are left-moving coherent states defined by
\begin{equation}
|\pm\lambda\rangle=\exp\left[  \pm i\int_{x_{1}}^{x_{2}}\lambda(x)\hat{\Pi
}_{+}\left(  x\right)  dx\right]  |0\rangle. \label{c1}%
\end{equation}
The two states $|\psi_{0}\rangle$ and $|\psi_{1}\rangle$ are non-orthogonal to
each other with $\langle\psi_{0}|\psi_{1}\rangle=\langle\lambda|-\lambda
\rangle\neq0$ because the measurement is not projective \cite{nc}. For
convenience, let us introduce the Fourier transformation of $\lambda(x)$ as
follows.
\begin{equation}
\tilde{\lambda}(\omega)=\int_{-\infty}^{\infty}\lambda(x)e^{i\omega x}dx,
\label{f1}%
\end{equation}
where the coefficient satisfies $\tilde{\lambda}^{\ast}(-\omega)=\tilde
{\lambda}(\omega)$. Note that the left-moving momentum operator $\hat{\Pi}%
_{+}\left(  x^{+}\right)  $ can be expanded as%

\begin{equation}
\hat{\Pi}_{+}\left(  x^{+}\right)  =-i\int_{0}^{\infty}d\omega\sqrt
{\frac{\omega}{\pi}}\left[  \hat{a}_{\omega}^{L}e^{-i\omega x^{+}}-\hat
{a}_{\omega}^{L\dag}e^{i\omega x^{+}}\right]  . \label{mode1}%
\end{equation}
Using Eqs. (\ref{c2}), (\ref{f1}), and (\ref{mode1}), we obtain the following
relation:
\begin{equation}
\exp\left[  -i\int_{x_{1}}^{x_{2}}\lambda(x)\hat{\Pi}_{+}\left(  x\right)
dx\right]  \hat{a}_{\omega}^{L}\exp\left[  i\int_{x_{1}}^{x_{2}}\lambda
(x)\hat{\Pi}_{+}\left(  x\right)  dx\right]  =\hat{a}_{\omega}^{L}-\sqrt
{\frac{\omega}{\pi}}\tilde{\lambda}(\omega). \label{502}%
\end{equation}
Therefore, the coherent state in Eq. (\ref{c1}) satisfies
\begin{equation}
\hat{a}_{\omega}^{L}|\pm\lambda\rangle=\mp\sqrt{\frac{\omega}{\pi}}%
\tilde{\lambda}(\omega)|\pm\lambda\rangle, \label{503}%
\end{equation}
which is derived from Eq. (\ref{502}) and $\hat{a}_{\omega}^{L}|0\rangle=0$.
Using $a_{\omega}^{L\dag}$, the state $|\pm\lambda\rangle$ can be expressed as%

\[
|\pm\lambda\rangle=\exp\left[  -\frac{1}{2}\int_{0}^{\infty}d\omega
\frac{\omega}{\pi}|\tilde{\lambda}(\omega)|^{2}\right]  \exp\left[  \mp
\int_{0}^{\infty}d\omega\sqrt{\frac{\omega}{\pi}}\tilde{\lambda}%
(\omega)a_{\omega}^{L\dag}\right]  |0\rangle.
\]
From this expression, we get the following inner products:%

\begin{equation}
\langle\lambda|-\lambda\rangle=\langle-\lambda|\lambda\rangle=\langle
0|2\lambda\rangle=\exp\left[  -2\int_{0}^{\infty}d\omega\frac{\omega}{\pi
}|\tilde{\lambda}(\omega)|^{2}\right]  . \label{800}%
\end{equation}

The expectational value of the Heisenberg operator of energy density
$\hat{\varepsilon}\left(  x,t\right)  $ for the\ post-measurement state in Eq.
(\ref{ps}) is given by%

\begin{align}
\langle\psi_{b}|\hat{\varepsilon}\left(  x,t\right)  |\psi_{b}\rangle &
=\frac{1}{2}\left[  \langle\lambda|\hat{\varepsilon}\left(  x,t\right)
|\lambda\rangle+\langle-\lambda|\hat{\varepsilon}\left(  x,t\right)
|-\lambda\rangle\right] \nonumber\\
&  -\left(  -1\right)  ^{b}\operatorname{Im}\langle\lambda|\hat{\varepsilon
}(x,t)|-\lambda\rangle. \label{1001}%
\end{align}
From Eq. (\ref{503}), it is easily checked that $\operatorname{Im}%
\langle\lambda|\hat{\varepsilon}(x,t)|-\lambda\rangle$ vanishes. In fact, we
can show%

\begin{align*}
&  \operatorname{Im}\langle\lambda|\hat{\varepsilon}\left(  x,t\right)
|-\lambda\rangle\\
&  =-\operatorname{Im}\left[  \int_{0}^{\infty}d\omega\frac{\omega}{\pi
}\left[  \tilde{\lambda}(\omega)e^{-i\omega x^{+}}+\tilde{\lambda}%
(\omega)^{\ast}e^{i\omega x^{+}}\right]  \right]  ^{2}\\
&  =0.
\end{align*}
Besides, Eqs. (\ref{504}), (\ref{503}), and (\ref{mode1}) yield the following
relations:
\[
\langle\lambda|\hat{\varepsilon}\left(  x,t\right)  |\lambda\rangle
=\langle-\lambda|\hat{\varepsilon}\left(  x,t\right)  |-\lambda\rangle=\left(
\partial_{+}\lambda(x^{+}\right)  )^{2}.
\]
Substituting the above equations into Eq. (\ref{1001}), we can conclude that
the measurement excites wave packets propagating to the horizon with an energy
density that does not depend on $b$ such that
\begin{equation}
\langle\psi_{b}|\hat{\varepsilon}\left(  x,t\right)  |\psi_{b}\rangle=\left(
\partial_{+}\lambda(x^{+}\right)  )^{2}.\label{ie1}%
\end{equation}
This result independent of the measurement result $b$ is striking as compared
to a particle detector model in the Rindler spacetime discussed by Unruh and
Wald. In the reference \cite{uw}, a particle detector coupled with a thermal
bath of a quantum field is analyzed which excites the field with positive
energy when the detector makes a transition and is observed in a excited
state. If the state of the detector is observed in the initial ground state,
we have different excitation of the field. Especially, in a causally
disconnected region, negative energy is induced by the detector. Thus amount
of energy induced by the detector in the model has explicit dependence of the
measurement results. However, in our POVM measurement model, the probe spin as
a detector excites the quantum field independent of the measurement result as
seen in Eq. (\ref{ie1}). \ Moreover, soon after the measurement, the energy
density for each post-measurement state vanishes except around the measurement
point. The amount of total average excitation energy is evaluated as
\begin{equation}
E_{A}=\int_{-\infty}^{\infty}\left(  \partial_{+}\lambda(x\right)
)^{2}dx.\label{700}%
\end{equation}
The energy flux falls into the horizon at $x^{-}=0$. \ Let us assume that at
$t=T,$ the left-moving excitation has already been eliminated from $\left[
x_{1},x_{2}\right]  $ and satisfies $\lambda(x+T)=0$ for $x\in$ $\left[
x_{1},x_{2}\right]  $. The average state at time $T$ is expressed as
\[
\hat{\rho}_{M}=\sum_{b}e^{-iT\hat{H}}\hat{M}_{b}|0\rangle\langle0|\hat{M}%
_{b}^{\dag}e^{iT\hat{H}}.
\]
It is worth noting that the state $\hat{\rho}_{M}$ is a strictly localized
state defined by Knight \cite{knight}, because $\hat{\rho}_{M}$ is locally the
same as $|0\rangle\langle0|$ at $t=T$ and satisfies $\operatorname*{Tr}\left[
\hat{\rho}_{M}\hat{\varepsilon}\left(  x\right)  \right]  =0$ for $x\in$
$\left[  x_{1},x_{2}\right]  $. Moreover, for each state $|\psi_{b}\rangle$,
$2n$-point functions of the left-moving operators $\hat{\Pi}_{+}$ with integer
$n$ are the same as those of $|0\rangle$ outside the horizon after time $T$.
Let us consider a $2n$-point function $\langle\psi_{b}|\hat{\Pi}_{+}%
(x_{1}+T)\cdots\hat{\Pi}_{+}(x_{2n}+T)|\psi_{b}\rangle$ and assume that all
$x_{k}$ satisfy $x_{k}>0$ and that they do not coincide with each other. Then,
using Eq. (\ref{ps}), the function is given by
\begin{align}
&  \langle\psi_{b}|\hat{\Pi}_{+}(x_{1}+T)\cdots\hat{\Pi}_{+}(x_{2n}%
+T)|\psi_{b}\rangle\nonumber\\
&  =\frac{1}{2}\left[
\begin{array}
[c]{c}%
\langle\lambda|\hat{\Pi}_{+}(x_{1}+T)\cdots\hat{\Pi}_{+}(x_{2n}+T)|\lambda
\rangle\\
+\langle-\lambda|\hat{\Pi}_{+}(x_{1}+T)\cdots\hat{\Pi}_{+}(x_{2n}%
+T)|-\lambda\rangle
\end{array}
\right]  \nonumber\\
&  -\left(  -1\right)  ^{b}\operatorname{Im}\langle\lambda|\hat{\Pi}_{+}%
(x_{1}+T)\cdots\hat{\Pi}_{+}(x_{2n}+T)|-\lambda\rangle.\label{505}%
\end{align}
Here, we have used the relation
\begin{align*}
&  \langle\lambda|\hat{\Pi}_{+}(x_{1}+T)\cdots\hat{\Pi}_{+}(x_{2n}%
+T)|-\lambda\rangle^{\ast}\\
&  =\langle-\lambda|\hat{\Pi}_{+}(x_{2n}+T)\cdots\hat{\Pi}_{+}(x_{1}%
+T)|\lambda\rangle\\
&  =\langle-\lambda|\hat{\Pi}_{+}(x_{1}+T)\cdots\hat{\Pi}_{+}(x_{2n}%
+T)|\lambda\rangle,
\end{align*}
which is ensured by the commutation relation%

\begin{equation}
\left[  \hat{\Pi}_{+}(x),~\hat{\Pi}_{+}(y)\right]  =2i\delta^{\prime}(x-y),
\label{508}%
\end{equation}
and the fact that all $x_{k}+T$ do not coincide with each other. Because
$\hat{f}$ is a free field, we can adopt the Wick's theorem for the composite
operator $\hat{\Pi}_{+}(x_{1}+T)\cdots\hat{\Pi}_{+}(x_{2n}+T)$ as
\begin{align}
&  \hat{\Pi}_{+}(x_{1}+T)\cdots\hat{\Pi}_{+}(x_{2n}+T)\nonumber\\
&  =\sum:\hat{\Pi}_{+}(x_{s_{1}}+T)\cdots\hat{\Pi}_{+}(x_{s_{2k}%
}+T):\nonumber\\
&  \times%
{\displaystyle\prod}
\langle0|\hat{\Pi}_{+}(x_{w_{1}}+T)\hat{\Pi}_{+}(x_{w_{2}}+T)|0\rangle,
\label{1002}%
\end{align}
where the contraction sum runs over only those contractions between operators
in different normal-ordered products. Then, using Eq. (\ref{1002}) and the
fact that $\lambda(x_{k}+T)=0$, we get
\begin{equation}
\langle\pm\lambda|\hat{\Pi}_{+}(x_{1}+T)\cdots\hat{\Pi}_{+}(x_{2n}%
+T)|\pm\lambda\rangle=\langle0|\hat{\Pi}_{+}(x_{1}+T)\cdots\hat{\Pi}%
_{+}(x_{2n}+T)|0\rangle. \label{1000}%
\end{equation}
This is because the normal ordered contributions $\langle\pm\lambda|:\hat{\Pi
}_{+}(x_{s_{1}}+T)\cdots\hat{\Pi}_{+}(x_{s_{2k}}+T):|\pm\lambda\rangle$
vanish:
\begin{align*}
\langle\pm\lambda|  &  :\hat{\Pi}_{+}(x_{s_{1}}+T)\cdots\hat{\Pi}%
_{+}(x_{s_{2k}}+T):|\pm\lambda\rangle\\
&  =2^{2k}\partial_{x}\lambda(x_{s_{1}}+T)\cdots\partial_{x}\lambda(x_{s_{2k}%
}+T)=0.
\end{align*}
Thus, $\langle\pm\lambda|\hat{\Pi}_{+}(x_{1}+T)\cdots\hat{\Pi}_{+}%
(x_{2n}+T)|\pm\lambda\rangle$ is reduced into a sum of products of $n$
two-point functions for the ground state $|0\rangle$ and equal to
$\langle0|\hat{\Pi}_{+}(x_{1}+T)\cdots\hat{\Pi}_{+}(x_{2n}+T)|0\rangle$:
\begin{align*}
&  \langle\pm\lambda|\hat{\Pi}_{+}(x_{1}+T)\cdots\hat{\Pi}_{+}(x_{2n}%
+T)|\pm\lambda\rangle\\
&  =\sum\langle0|\hat{\Pi}_{+}(x_{w_{1}}+T)\hat{\Pi}_{+}(x_{w_{2}}%
+T)|0\rangle\cdots\langle0|\hat{\Pi}_{+}(x_{w_{2n-1}}+T)\hat{\Pi}%
_{+}(x_{w_{2n}}+T)|0\rangle\\
&  =\langle0|\hat{\Pi}_{+}(x_{1}+T)\cdots\hat{\Pi}_{+}(x_{2n}+T)|0\rangle.
\end{align*}
By use of Eq. (\ref{1002}), it can be also proven that
\begin{equation}
\operatorname{Im}\langle\lambda|\hat{\Pi}_{+}(x_{1}+T)\cdots\hat{\Pi}%
_{+}(x_{2n}+T)|-\lambda\rangle=0. \label{507}%
\end{equation}
This is because the following relation holds:
\begin{align*}
\langle\lambda|  &  :\hat{\Pi}_{+}(x_{s_{1}}+T)\cdots\hat{\Pi}_{+}(x_{s_{2k}%
}+T):|-\lambda\rangle\\
&  =\left(  -1\right)  ^{k}Q(x_{s_{1}}+T)\cdots Q(x_{s_{2k}}+T)\langle
\lambda|-\lambda\rangle,
\end{align*}
where $Q(x+T)$ is a real function defined by
\[
Q(x+T)=-\int_{0}^{\infty}d\omega\frac{\omega}{\pi}\left[  \tilde{\lambda
}(\omega)e^{-i\omega\left(  x+T\right)  }+\tilde{\lambda}(\omega)^{\ast
}e^{i\omega\left(  x+T\right)  }\right]  ,
\]
and $\langle\lambda|-\lambda\rangle$ is also real as seen in Eq. (\ref{800}).
Moreover, taking into account the fact that all $x_{k}$ do not coincide with
each other, it can be observed that the two-point functions $\langle0|\hat
{\Pi}_{+}(x_{w_{1}}+T)\hat{\Pi}_{+}(x_{w_{2}}+T)|0\rangle$ are real. Thus, by
using the above results and Eq. (\ref{1002}), we obtain Eq. (\ref{507}). By
substitution of Eqs. (\ref{1000}) and (\ref{507}) into Eq. (\ref{505}), it is
proven that the $2n$-point function in the outside region after time $T$ is
the same as that of the vacuum state:%

\begin{align*}
&  \langle\psi_{b}|\hat{\Pi}_{+}(x_{1}+T)\cdots\hat{\Pi}_{+}(x_{2n}%
+T)|\psi_{b}\rangle\\
&  =\langle0|\hat{\Pi}_{+}(x_{1}+T)\cdots\hat{\Pi}_{+}(x_{2n}+T)|0\rangle.
\end{align*}
It is worth noting that many-point functions of the energy density operator
$\hat{\varepsilon}(x,T)$ can be obtained from these $2n$-point functions with
the point-splitting regularization scheme \cite{bd}. Besides the left-moving
mode, the many-point functions of the right-moving mode for $|\psi_{b}\rangle$
are the same as those for $|0\rangle$. Therefore, after the wave packets are
absorbed by the black holes, the fluctuations in energy density for the state
$|\psi_{b}\rangle$ are the same as those of the vacuum\ $|0\rangle$:%

\[
\langle\psi_{b}|\hat{\varepsilon}(x_{1},T)\cdots\hat{\varepsilon}%
(x_{n},T)|\psi_{b}\rangle=\langle0|\hat{\varepsilon}(x_{1},T)\cdots
\hat{\varepsilon}(x_{n},T)|0\rangle.
\]
Because the massless field can affect the spacetimes of black holes only via
energy density, the difference in the measurement result $b$ yields no
difference in black-hole spacetimes. The difference in $b$ appears only in
odd-number many-point functions of $\hat{\Pi}_{+}$. For example, the one-point
function of $\hat{\Pi}_{+}$ for $|\psi_{b}\rangle$ is dependent on $b$ and is
given by%
\begin{equation}
\langle\psi_{b}|\hat{\Pi}_{+}(x+T)|\psi_{b}\rangle=(-1)^{b}\int_{0}^{\infty
}d\omega\frac{\omega}{\pi}\left[  \tilde{\lambda}(\omega)e^{-i\omega\left(
x+T\right)  }+\tilde{\lambda}(\omega)^{\ast}e^{i\omega\left(  x+T\right)
}\right]  . \label{778}%
\end{equation}

We recall here that we have $N$ copies of the field $\hat{f}$. Thus, the
measurement output bit ($b_{n}=0$ or $1$) is obtained for each field $\hat
{f}_{n}$ and the total data of the measurements are recorded as $D_{k}%
=(b_{1}b_{2}\cdots b_{N})$ with $k=1\sim2^{N}$. If we take a large $N$,
quantum fluctuation in the total energy density of the $N\,$ fields is
severely suppressed due to the law of large numbers, and the back reaction of
the quantum fields to the black hole can be treated by a classical gravity
equation of the metric. In the equation, the average energy flux as a source
term is given by $N\left(  \partial_{+}\lambda(x^{+}\right)  )^{2}$. The
falling flux yields a horizon shift of $\delta X_{H}^{-}$ independent of $b$
in the $X^{\pm}$ coordinates as $\delta X_{H}^{-}=-\frac{N}{2\lambda^{2}}%
\sqrt{\frac{M}{\lambda}}\int_{-\infty}^{\infty}\left(  \partial_{+}%
\lambda(x^{+}\right)  )^{2}dx^{+}$.

Figure 2 shows the horizon shift. The initial horizon ($X^{-}=0$) is denoted
by $H_{I}$ in the figure. The new horizon after the shift is denoted by
$H_{F}$. It should be noted that a spacetime region becomes static soon after
the flux passes through the region. If we do nothing more, $H_{F}$ is the
final event horizon. However, as seen below, if the unitary operation
dependent on $D_{k}$ is performed, negative energy flux is created and the
horizon shifts again. The receding horizon is denoted by $H_{QET}$. The
argument does not change at all if the falling energy flux interacts with
collapsing matter within the black hole. In Figure 2, right-moving matter
represented by dashed lines\ collides with the falling positive-energy flux
inside the horizon.

At $t=T$, we perform a unitary operation on the quantum field $\hat{f}$; the
unitary operation is dependent on $b$ and is given by%

\begin{equation}
\hat{U}_{b}=\exp\left[  i\theta(-1)^{b}\int_{-\infty}^{\infty}p(x)\hat{\Pi
}_{+}\left(  x\right)  dx\right]  , \label{509}%
\end{equation}
where $\theta$ is a real parameter fixed below and $p(x)$ is a real function
with the same support of $\lambda(x)$, $\left[  x_{1},x_{2}\right]  $. We do
not require any classical communication phase for this black-hole QET because
the operation \ in Eq. (\ref{509}) is performed in the same region $\left[
x_{1},x_{2}\right]  $ as that where the measurement is executed. After the
operation, the average state of the field $\hat{f}$ is given by%

\[
\hat{\rho}_{F}=\sum_{b=0,1}\hat{U}_{b}e^{-iT\hat{H}}\hat{M}_{b}|0\rangle
\langle0|\hat{M}_{b}^{\dag}e^{iT\hat{H}}\hat{U}_{b}^{\dag}.
\]

Let us introduce an energy operator localized around the region $\left[
x_{1},x_{2}\right]  ~$such that $\hat{H}_{B}=\int_{-\infty}^{\infty}w\left(
x\right)  \hat{\varepsilon}\left(  x\right)  dx$. Here, $w\left(  x\right)  $
is a real window function with $w(x)=1$ for $x\in\left[  x_{1},x_{2}\right]  $
and it rapidly decreases outside the region. The average amount of energy
around the region is evaluated as%

\begin{align*}
E_{B}  &  =\operatorname*{Tr}\left[  \hat{\rho}_{F}\hat{H}_{B}\right] \\
&  =\sum_{b=0,1}\langle0|\hat{M}_{b}^{\dag}e^{iT\hat{H}}\hat{U}_{b}^{\dag}%
\hat{H}_{B}\hat{U}_{b}e^{-iT\hat{H}}\hat{M}_{b}|0\rangle.
\end{align*}
In order to simplify the expression of $E_{B}$, a useful relation is available:%

\[
\hat{U}_{b}^{\dag}\hat{H}_{B}\hat{U}_{b}=\hat{H}_{B}-\theta(-1)^{b}\int
_{x_{1}}^{x_{2}}\partial_{x}p(x)\hat{\Pi}_{+}(x)dx+\theta^{2}\int_{x_{1}%
}^{x_{2}}\left(  \partial_{x}p(x)\right)  ^{2}dx.
\]
This is derived from Eqs. (\ref{508}) and (\ref{509}). Thus, $E_{B}$ is
rewritten as%

\begin{align}
E_{B}  &  =\sum_{b}\langle0|\hat{M}_{b}^{\dag}e^{iT\hat{H}}\hat{H}%
_{B}e^{-iT\hat{H}}\hat{M}_{b}|0\rangle\nonumber\\
&  -\theta\sum_{b}(-1)^{b}\langle0|\hat{M}_{b}^{\dag}\int_{x_{1}}^{x_{2}%
}\partial_{x}p(x)\hat{\Pi}_{+}(x+T)dx\hat{M}_{b}|0\rangle\nonumber\\
&  +\theta^{2}\int\left(  \partial_{x}p(x)\right)  ^{2}dx. \label{512}%
\end{align}
Because we assume that $T>\left\vert x_{2}-x_{1}\right\vert $, the following
two relations can be directly proven:
\begin{equation}
\left[  e^{iT\hat{H}}\hat{H}_{B}e^{-iT\hat{H}},~\hat{M}_{b}\right]  =0,
\label{510}%
\end{equation}%
\begin{equation}
\left[  \int_{x_{1}}^{x_{2}}\partial_{x}p(x)\hat{\Pi}_{+}(x+T)dx,~\hat{M}%
_{b}\right]  =0. \label{511}%
\end{equation}
From Eqs. (\ref{401}) and (\ref{402}), we are able to check the following relations:%

\begin{equation}
\sum_{b}\hat{M}_{b}^{\dag}\hat{M}_{b}=1, \label{403}%
\end{equation}

\begin{equation}
\sum_{b}(-1)^{b}\hat{M}_{b}^{\dag}\hat{M}_{b}=\cos\left(  2\hat{G}_{+}\right)
. \label{404}%
\end{equation}
Substituting Eqs. (\ref{510})--(\ref{404}) into Eq. (\ref{512}), we obtain the
following expression of $E_{B}$:%
\begin{align*}
E_{B}  &  =\langle0|e^{iT\hat{H}}\hat{H}_{B}e^{-iT\hat{H}}|0\rangle\\
&  -\theta\int_{x_{1}}^{x_{2}}dx\partial_{x}p(x)\langle0|\hat{\Pi}%
_{+}(x+T)\cos\left(  2\hat{G}_{+}\right)  |0\rangle\\
&  +\theta^{2}\int\left(  \partial_{x}p(x)\right)  ^{2}dx.
\end{align*}
Note that%

\[
\langle0|e^{iT\hat{H}}\hat{H}_{B}e^{-iT\hat{H}}|0\rangle=\langle0|\hat{H}%
_{B}|0\rangle=\int_{-\infty}^{\infty}w\left(  x\right)  \langle0|\hat
{\varepsilon}\left(  x\right)  |0\rangle dx=0.
\]
Therefore, the average energy $E_{B}=\operatorname*{Tr}\left[  \hat{\rho}%
_{F}\hat{H}_{B}\right]  $ after the operation is computed as
\[
E_{B}=-\theta\eta+\theta^{2}\xi,
\]
where $\xi=\int_{x_{1}}^{x_{2}}\left(  \partial_{x}p(x)\right)  ^{2}dx$ and%
\begin{equation}
\eta=\int dx\partial_{x}p(x)\langle0|\hat{\Pi}_{+}(x+T)\cos\left(  2\hat
{G}_{+}\right)  |0\rangle. \label{603}%
\end{equation}
It is possible to simplify this expression of $\eta$. Using Eqs. (\ref{mode1})
and (\ref{503}) and the relation given by%

\[
\cos\left(  2\hat{G}_{+}\right)  |0\rangle=\frac{i}{2}\left(  |+2\lambda
\rangle-|-2\lambda\rangle\right)  ,
\]
we obtain the following relation:%

\[
\langle0|\hat{\Pi}_{+}(x+T)\cos\left(  2\hat{G}_{+}\right)  |0\rangle
=-\frac{2}{\pi}\langle0|2\lambda\rangle\int_{0}^{\infty}d\omega\omega
\tilde{\lambda}(\omega)e^{-i\omega\left(  x+T\right)  }.
\]
Substituting Eq. (\ref{f1}) into the above relation, we obtain
\begin{align}
&  \langle0|\hat{\Pi}_{+}(x+T)\cos\left(  2\hat{G}_{+}\right)  |0\rangle
\nonumber\\
&  =-\frac{2}{\pi}\langle0|2\lambda\rangle\int_{0}^{\infty}dy\lambda
(y)\frac{1}{\left(  x-y+T\right)  ^{2}}. \label{606}%
\end{align}
By substituting Eq. (\ref{606}) into Eq. (\ref{603}), we get the final
expression of $\eta\,$\ as%

\begin{equation}
\eta=-\frac{4}{\pi}\left\vert \langle0|2\lambda\rangle\right\vert \int_{x_{1}%
}^{x_{2}}\int_{x_{1}}^{x_{2}}p(x)\frac{1}{\left(  x-y+T\right)  ^{3}}%
\lambda(y)dxdy. \label{777}%
\end{equation}
By fixing the parameter $\theta$ such that
\[
\theta=\frac{\eta}{2\xi}%
\]
so as to minimize $E_{B}$, it is proven that the average energy around
$\left[  x_{1},x_{2}\right]  $ of $\hat{f}$ takes a negative value, that is,%

\begin{equation}
E_{B}=-\frac{\eta^{2}}{4\xi}<0. \label{701}%
\end{equation}
By virtue of local energy conservation, this result implies that the unitary
operation extracts positive energy $+|E_{B}|$ from the field to the external
systems. Taking into account the existence of $N$ fields, the total amount of
extracted energy is $N|E_{B}|$. The measurement step for $D_{k}=(101\cdots0)$
is shown in Figure 3. In Figure 4, the extraction step is depicted. This
process is analogous to the standard Hawking process of black holes, that is,
pair creation of particles with positive and negative energy outside the
horizon \cite{h}. Just like in the Hawking process, the created negative
energy $-N|E_{B}|$ of the fields falls into the horizon and decreases the
black hole mass, accompanying the second shift of the horizon ($H_{F}%
\rightarrow H_{QET}$), as shown in Figure 2. For each field, the sum of the
positive energy generated by the measurement and the negative energy generated
by the operation is always positive, that is, $E_{A}+E_{B}>0$. This means that
$E_{A}>\left\vert E_{B}\right\vert $, and thus only a part of the absorbed
positive energy $E_{A}$ can be retrieved from a black hole by this QET process.

When we consider higher-dimensional spherical black holes, essentially the
same results are obtained. The POVM measurement and operation for energy
extraction are performed non-locally over a spherical-shell region surrounding
an event horizon with a fixed radius.

\bigskip

\section{Summary and Discussion}

\ \newline

In this paper, a new quantum mechanical method to extract energy from black
holes with contracting horizons is proposed. This quantum mechanical method is
based on a gedanken experiment on quantum energy teleportation (QET), which
has been recently proposed in quantum information theory. Near-horizon regions
of large-mass black holes, a POVM measurement defined by Eqs. (\ref{401}) and
(\ref{402}) is performed for quantum fluctuation of $N$ massless scalar fields
in the Hartle-Hawking state, which is effectively described by the Minkowski
vacuum state $|0\rangle$. Then, we obtain $N$-bit information about the
fluctuation denoted by $D_{k}$. During the measurement, wave packets of the
fields with positive energy given by Eq. (\ref{700}) form and then fall into
the black hole. The amount of excitation energy is independent of the
measurement results. After absorption of the wave packets and increase in the
black hole mass, a $D_{k}$-dependent local operation of the $N~$fields given
by Eq. (\ref{509}) is performed outside the horizon. Then, accompanying the
extraction of positive energy from the fluctuation by the operation, wave
packets of the fields with negative energy given by Eq. (\ref{701}) form and
then fall into the black hole. Because both the mass and the entropy of the
black hole decrease due to absorption of the negative flux, this QET process
is analogous to the Hawking process generated by pair creation of excitations
with positive and negative energy, although the energy extracted by QET does
not show the thermal properties.

In general, black holes swallow various kinds of information from falling
matter, which increases the area of the horizon. Because falling matter with
the same mass, charge, and angular momentum generates the same geometry in
classical theory, classical black holes forget the information after the
shifted horizon stabilizes. Meanwhile, many physicists believe, respecting the
unitarity of theory, that quantum black holes do not forget the information
and that the detailed memory is stored in some quantum mechanical ways.
However, though many efforts to understand the quantum memory storage
mechanism have been made so far, the complete resolution remains elusive. Even
in the case of large-mass static black holes, the question as to where the
information is stored in the black-hole spacetimes has not been completely
resolved. The information might be stored inside the horizons, on the
horizons, or even outside the horizons. This is the so-called black hole
entropy problem. From the results of this paper, we can conclude that some
memories of absorbed quantum matter remains outside the horizon for a while,
even after the shifted horizons are completely settled. The reason is
following. If the unitary operation in Eq. (\ref{509}) corresponding to a
different $N$-bit sequence $D_{k^{\prime}\neq k}$ is performed after the
black-hole absorption of wave packets with information $D_{k}$, the amount of
energy extracted by QET always decreases. Any quantum operation $U_{b}$ with a
wrong bit number $b$ does not generate negative energy; instead, it generates
positive energy of the field. Hence, this execution of $U_{b}$ needs work. The
amount of energy is $E_{B}^{\prime}=3\eta^{2}/4\xi(>0)$. Figure 5 shows a
schematic diagram for the case of a wrong $N$-bit sequence $D_{k^{\prime}%
}=(011\cdots1)$. Thus, if the wrong bit number per $N$ exceeds $1/4$, there is
no energy gain from the black hole on average by the QET process for the $N$
fields. In an extreme case, a random choice of $U_{b}$ yields $\eta^{2}/4\xi$
energy input to the field on average. The black hole returns the maximum
amount of energy by QET if the unitary operation obeying the correct $D_{k}$
is performed on the $N$ fields. Therefore, the black hole remembers which
$D_{k}$ state was absorbed. This implies that the black holes absorbing
quantum states with different $D_{k}$ are physically different even though
their black-hole geometries are the same outside the horizons. In this case,
the information of $D_{k}$ is imprinted in the quantum fluctuation of the
fields outside the horizon. In fact, in the region out of the horizon, the
one-point function of $\hat{\Pi}_{+}$ depend on the measurement result $D_{k}$
as seen in Eq. (\ref{778}). Though this observation is of much interest, these
memories may be, in a precise sense, not black hole hairs contributing to the
black-hole entropy. This is because the memories are not stored eternally, but
may decay in time $T$ as suggested by Eq. (\ref{777}). This loss of memories
over time suggests that the black-hole QET can be used as a probe for the
thermal relaxation process of the black hole to a true quantum thermal
equilibrium state.

Before closing this discussion, I would like to add a comment about the Unruh
effect in the Rindler spacetime \cite{u} \cite{uw}. If we consider a detector
coupled with quantum fluctuation of fields and uniformly accelerated in the
vacuum, we observe thermal excitation in the detector as the Unruh effect.
This effect is artificially caused by measurements and not a spontaneous
effect. The artificially induced effect is similar to the QET method described
in this paper. However, the QET method is completely different from the Unruh
effect. The detector of our POVM measurement is coupled with quantum
fluctuation for a very short duration, though detectors used for observation
of the Unruh effect must be coupled with fluctuation over a very long period
of time and should be simultaneously accelerated uniformly. The detector for
QET is not accelerated at all. Moreover, the measurement results read out of
the detector are essential for QET. We give a quantum feedback to the field
based on the measurement result. Because of the dependence on the measurement
result, the local operation given in Eq. (\ref{509}) can generate negative
energy on average of the field with extraction of positive energy outside the
horizon. However, the measurement result in the Unruh effect is not reused to
give any feedback to the field. This is a quite different point from our
model. In the reference \cite{uw}, Unruh and Wald also discussed a correlation
of response of a particle detector with energy generated in a causally
disconnected region from the detector of the Rindler spacetime. This causally
disconnected region corresponds to inside an eternal black hole when the
detector is located outside the horizon. Interestingly, positive energy
density is induced in the causally disconnected region when the particle
detector outside the \ horizon is observed in the excited state. When the
detector is observed in the ground state, negative energy is induced in the
region such that the average energy density in the causally disconnected
region remains zero which is equal to the vacuum-state value. However, it
should be stressed that the energy induced by the detector is located inside
the horizon and cannot be used for a person who stays outside the horizon.
Moreover, the average energy (mass) of the black hole does not change at all
after the outside-horizon measurement. This feature is completely different
from that of our QET model. In the QET model, the person outside the horizon
can effectively get positive energy from the black hole via quantum
fluctuation of the field outside the horizon.~\ The negative energy given by
Eq. (\ref{701}) is generated outside the horizon by the measurement and falls
into the black hole so as to decrease the black hole energy.

\bigskip

\textbf{Acknowledgments}\newline

I would like to thank Y. Itoh, M. Ozawa, and A. Hosoya for useful discussions.
This research is partially supported by the Ministry of Education, Science,
Sports and Culture of Japan, No. 21244007.

\bigskip

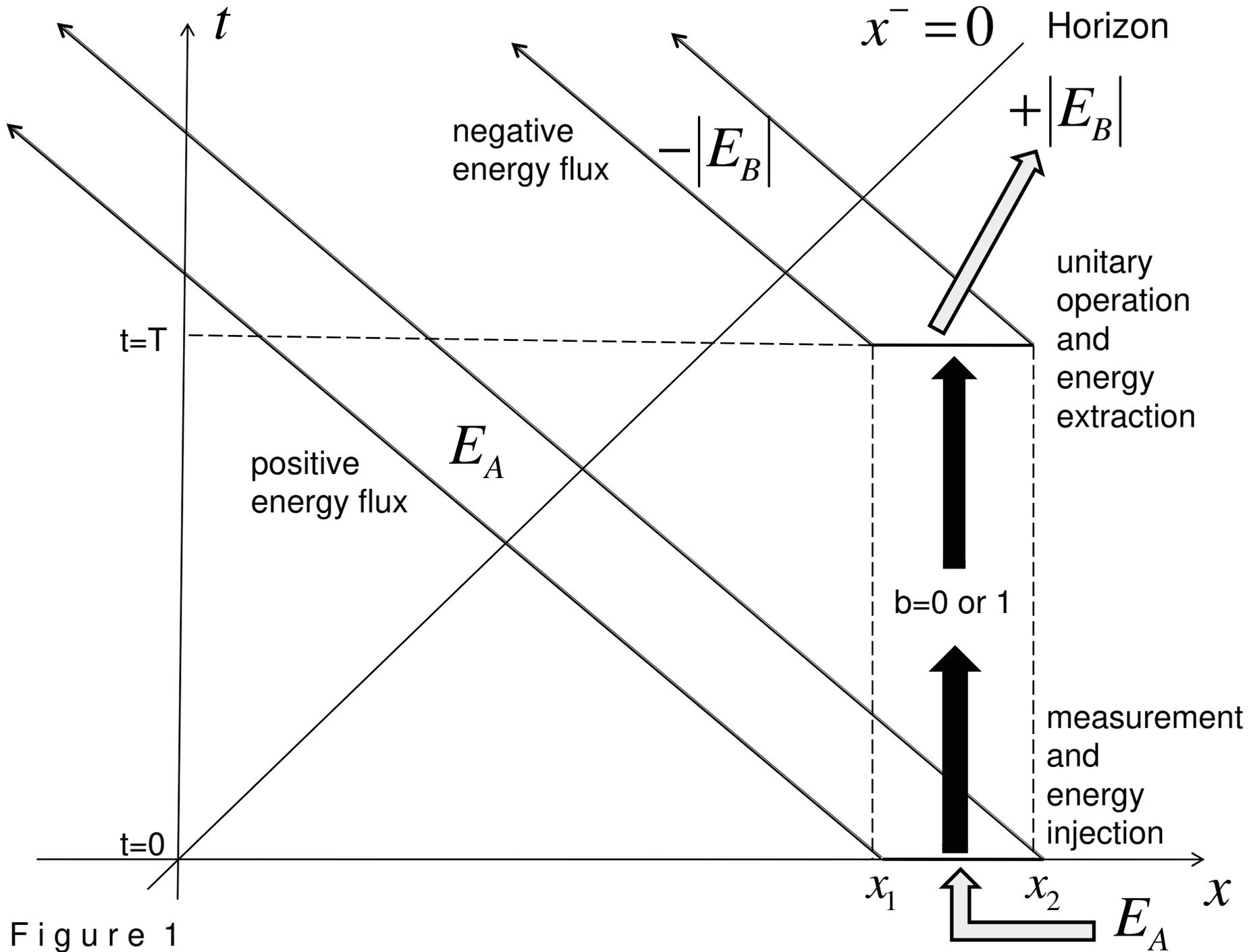
Figure 1: Spacetime diagram of the QET process in the near-horizon region. The
future horizon stays at $x^{-}=0$. At $t=0$, quantum fluctuation of $\hat{f}$
$\ $in the spatial region $[x_{1},x_{2}]$ is indirectly measured by a spin
probe. A wave packet with positive energy $E_{A}$ is generated by the
measurement device; thereafter, it falls into the future horizon. At $t=T$,
the unitary operation $U_{b}$, which is dependent on the measurement result
$b$, is performed; this operation generates a wave packet with negative energy
$-|E_{B}|$, which falls into the horizon by extracting positive energy from
the field.

\bigskip

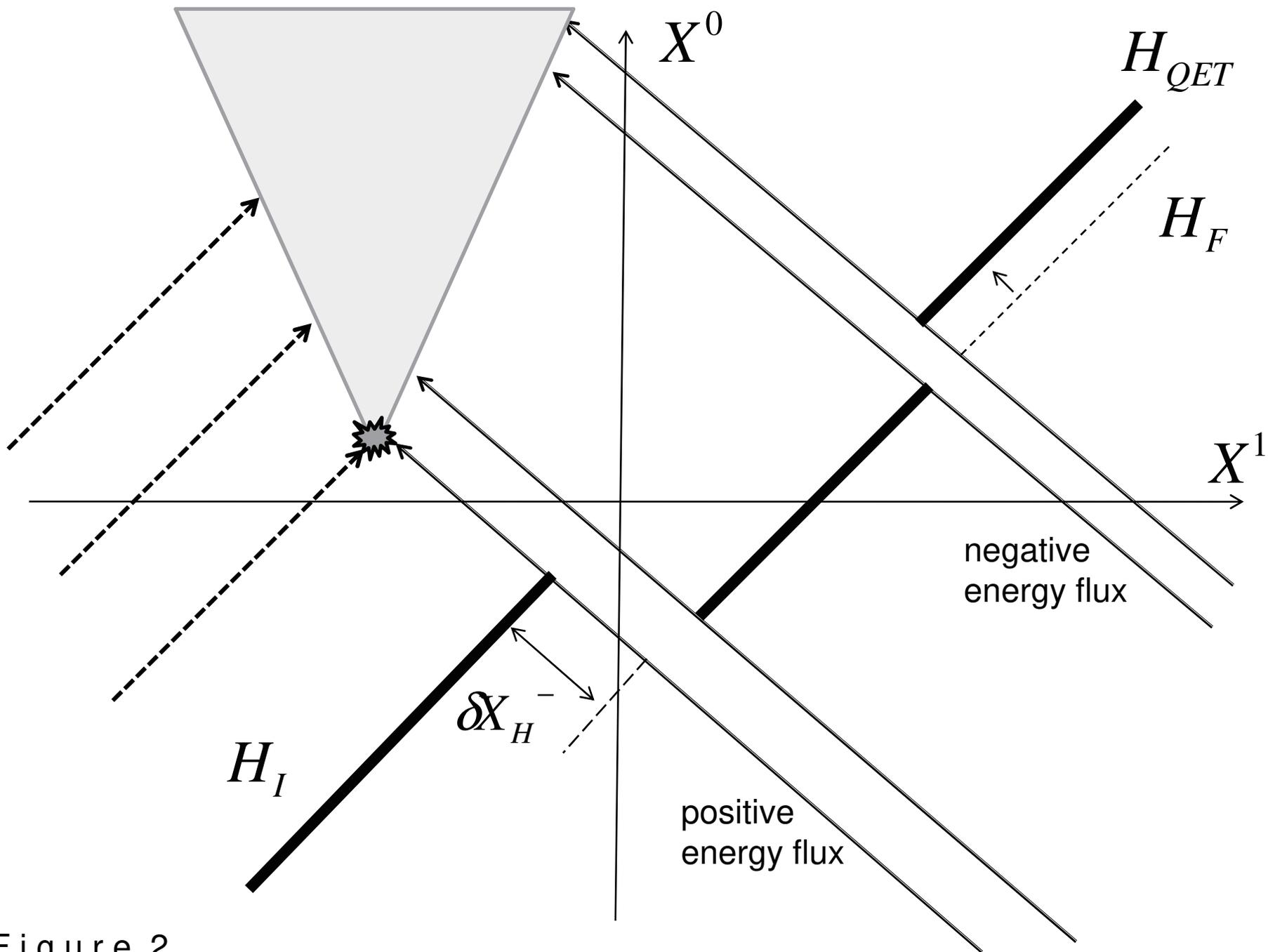
Figure 2: Spacetime diagram of the horizon shift in the $X$ coordinates. The
initial horizon $H_{I}$ stays at $X^{-}=0$. By absorbing the positive energy
flux generated by the measurement, the horizon moves outwards. $H_{F}$ denotes
the shifted horizon. After the absorption of negative energy flux generated by
$U_{b}$, the horizon recedes. The shifted horizon is denoted by $H_{QET}$. The
positions of the horizons are fixed soon after the wave packets are swallowed.
The argument does not change at all if the falling energy flux interacts with
matter within the black hole. In the diagram, right-moving matter, which is
represented by dashed lines, collides with the falling positive-energy flux
inside the horizon.

\bigskip

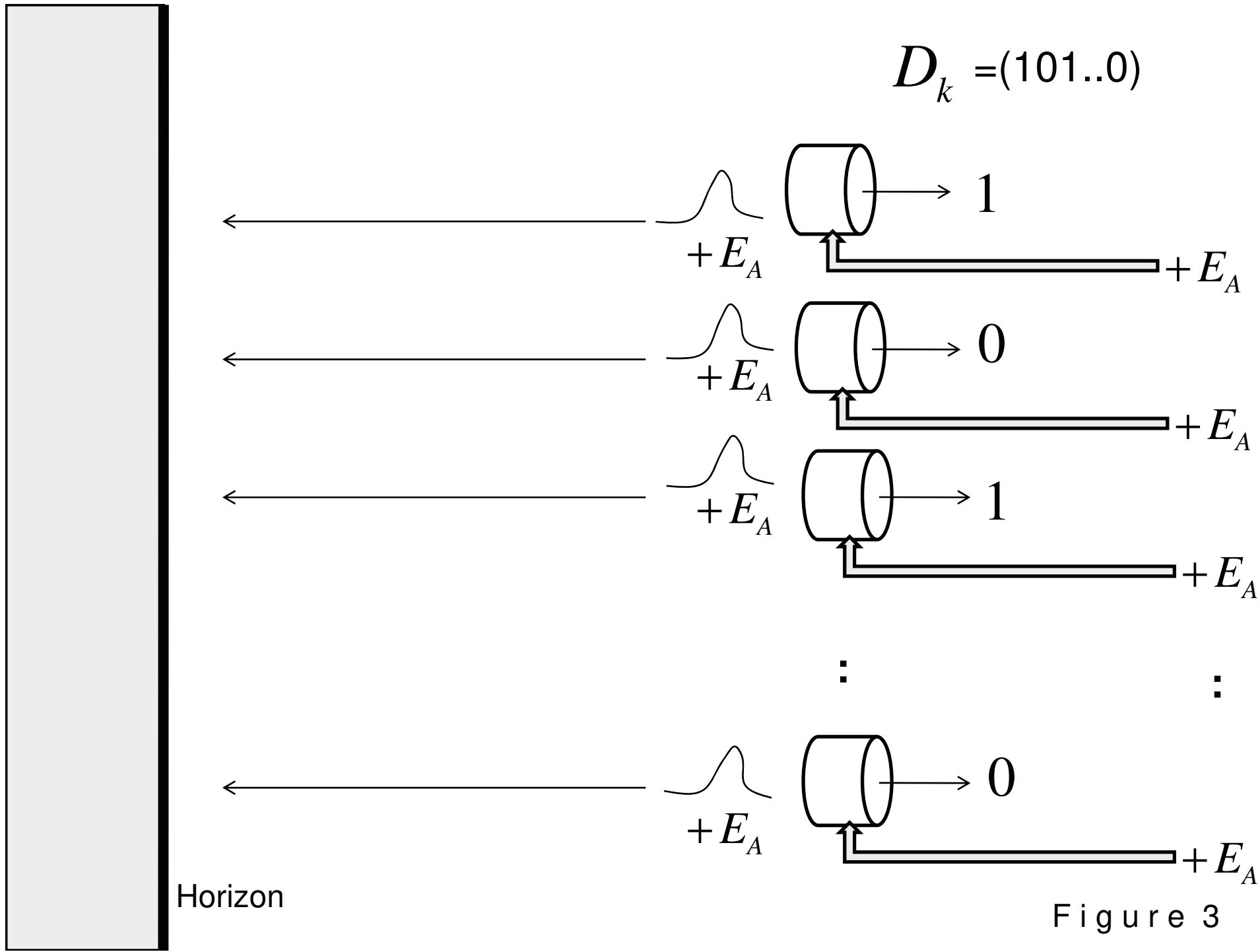
Figure 3: Schematic diagram of the measurement process of QET. The $N$
cylinders in the figure represent the measurement devices used to detect the
quantum fluctuation of the $N$ fields. In this figure, the $N$-bit output of
the devices is $D_{k}=(101\cdots0)$. Each detector creates a wave packet with
positive energy $E_{A}$ that is independent of $b$. The wave packets are
absorbed by the horizon, which is represented by the shaded region to the left.

\bigskip

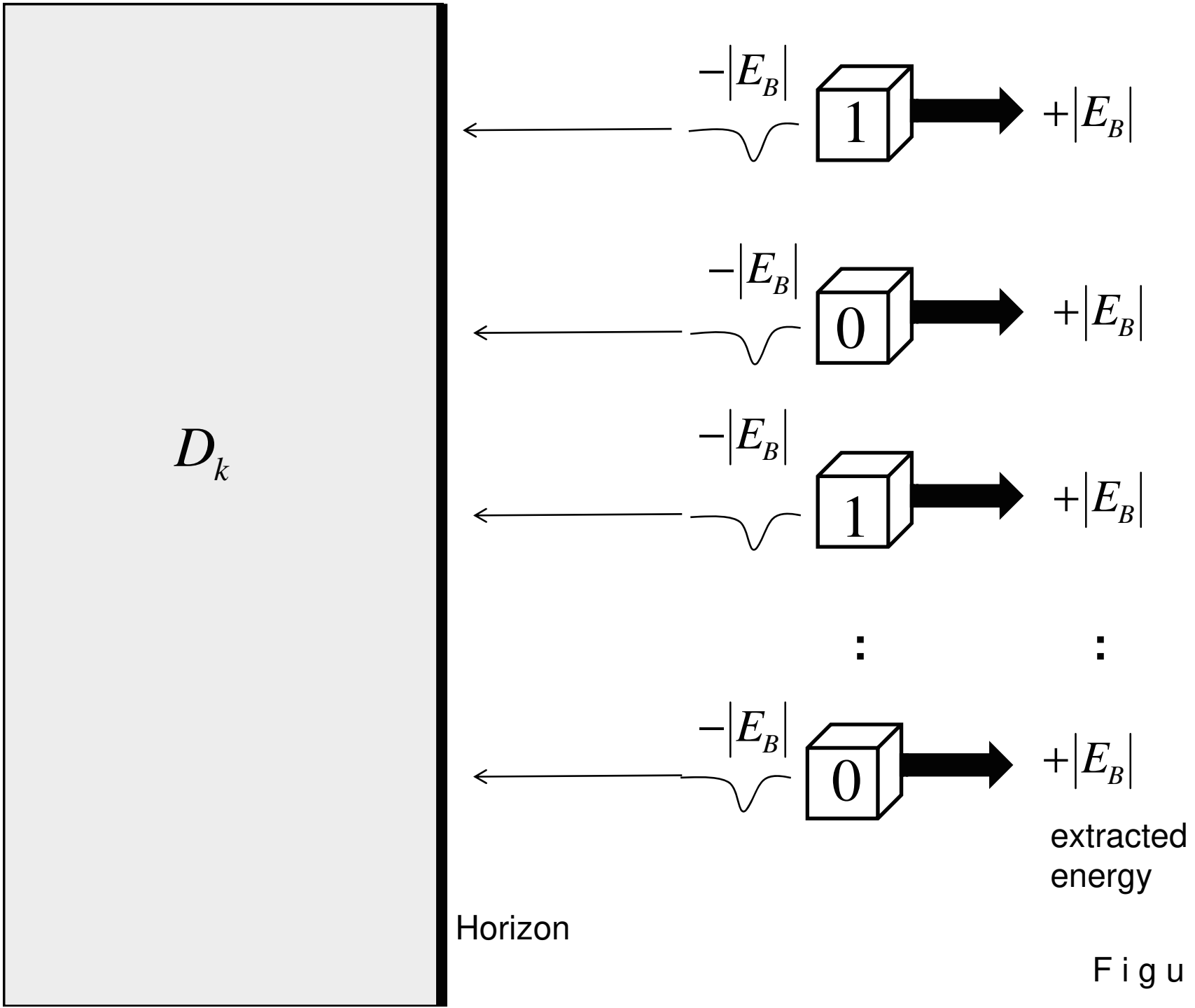
Figure 4: Schematic diagram of a unitary operation process that is dependent
on the measurement result. The $N$ cubes in the figure represent the operation
devices of the unitary operation for the $N$ fields. After the operation, a
wave packet with negative energy $-|E_{B}|$ is emitted from each device and is
absorbed by the horizon. Before this, the horizon has already absorbed the
wave packets corresponding to $D_{k}=(101\cdots0)$ that are generated by the
measurement. Local energy conservation assures extraction of positive energy
$+|E_{B}|$ from each field by the device.

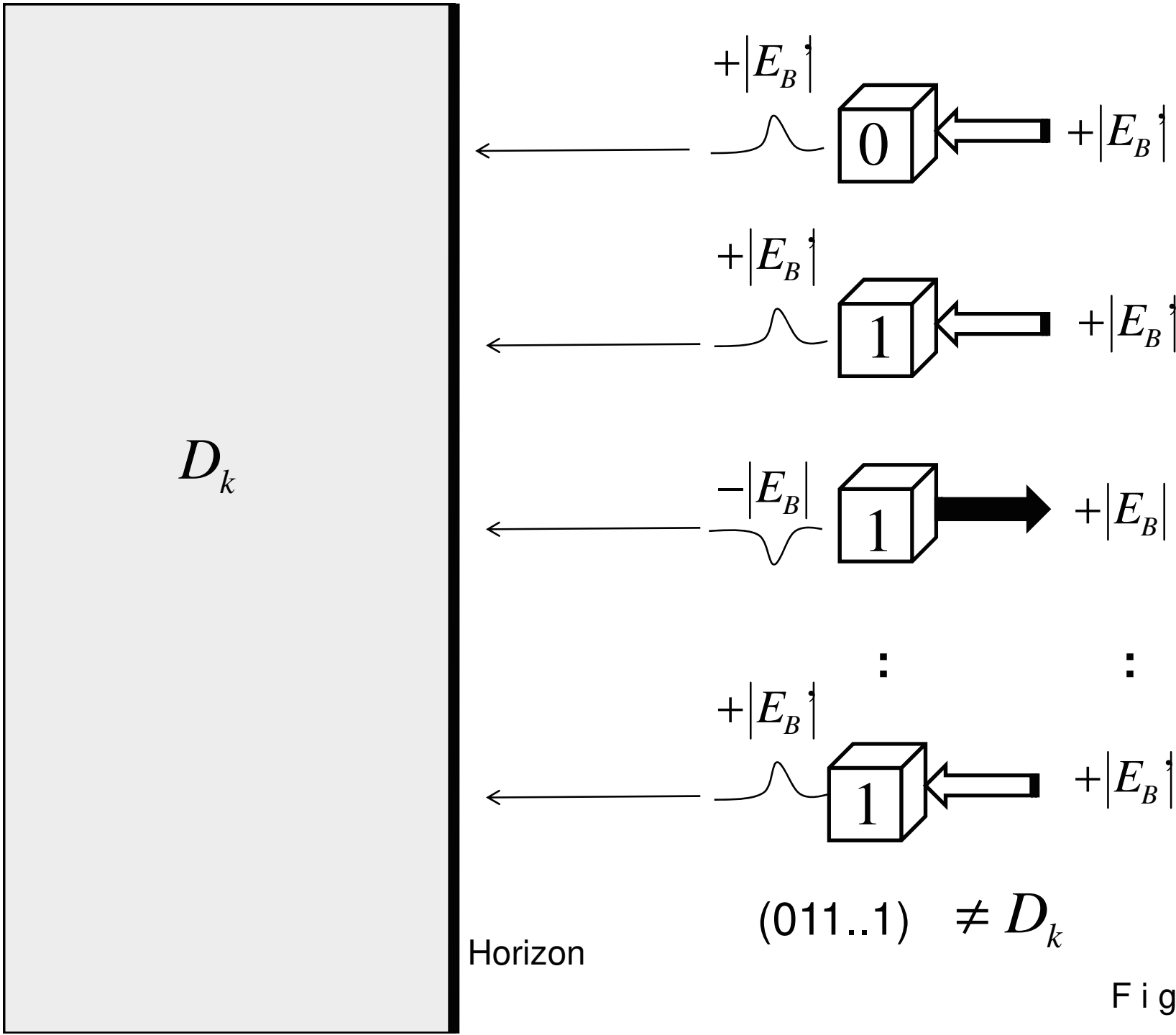
Figure 5: Schematic diagram of a wrong operation on the quantum fields. In the
figure, a unitary operation corresponding to an $N$-bit data $(011\cdots1)$
that is different from the correct data $(101\cdots0)$ is performed. Devices
with wrong bit numbers do not generate negative energy flux, but generate
positive energy flux with $+\left\vert E_{B}^{\prime}\right\vert =3\eta
^{2}/4\xi$. When the wrong bit number per $N$ exceeds $1/4$, there is no
energy gain from the black holes on average by the QET process.

\end{document}